\documentclass[preprint]{aastex}
\usepackage{amssymb,amsmath}

\renewcommand{\thefootnote}{\fnsymbol{footnote}}

\begin{document}

\title{Dramatic Evolution of the Disk-Shaped Secondary in the Orion Trapezium Star $\theta^{1}$ Ori B$_1$ (BM Ori): MOST Satellite Observations}

\author{Diana Windemuth, William Herbst, Evan Tingle\altaffilmark{1}, Rachel Fuechsl\altaffilmark{2}, and Roy Kilgard}
\affil{Astronomy Department, Wesleyan University, Middletown, CT
06459}
\email{wherbst@wesleyan.edu}

\author{Melanie Pinette\altaffilmark{3}}
\affil{Department of Physics and Astronomy, Bowdoin College}

\author{Matthew Templeton and Arne Henden}
\affil{American Association of Variable Star Observers, Cambridge, MA 02138}

\altaffiltext{1} {Current Address: High Energy Astrophysics Division, Harvard-Smithsonian Center for Astrophysics, 60 Garden St., Cambridge, MA 02138}
\altaffiltext{2} {Current Address: McDonald Observatory, The University of Texas at Austin, Austin, TX 78712}
\altaffiltext{3} {Current Address: Engineering Management Program, Duke University, Durham, NC 27710}

\begin{abstract}

The eclipsing binary $\theta^{1}$ Orionis B$_1$, variable star designation BM Ori, is the faintest of the four well-known {\it Trapezium} stars at the heart of the Orion Nebula. The primary is a B3 star ($\sim$6 M$_\odot$) but the nature of the secondary ($\sim$2 M$_\odot$) has long been mysterious, since the duration and shape of primary eclipse are inappropriate for any sort of ordinary star. Here we report nearly continuous photometric observations obtained with the MOST satellite over $\sim$4 cycles of the 6.47 d binary period. The light curve is of unprecedented quality, revealing a deep, symmetric primary eclipse as well as a clear reflection effect and secondary eclipse. In addition, there are other small disturbances, some of which repeat at the same phase over the four cycles monitored. The shape of the primary light curve has clearly evolved significantly over the past 40 years. While its overall duration and depth have remained roughly constant, the slopes of the descent and ascent phases are significantly shallower now than in the past and its distinctive flat-bottomed ``pseudo-totality" is much less obvious or even absent in the most recent data. We further demonstrate that the primary eclipse was detected at X-ray wavelengths during the {\it Chandra} Orion Ultradeep Project (COUP) study. The light curve continues to be well modeled by a self-luminous and reflective disk-shaped object seen nearly edge-on orbiting the B3 primary. The dramatic change in shape over four decades is modeled as an opacity variation in a tenuous outer envelope or disk of the secondary object. We presume that the secondary is an extremely young protostar at an earlier evolutionary phase than can be commonly observed elsewhere in the Galaxy and that the opacity variations observed are related to its digestion of some accreted matter over the last 50 - 100 years. Indeed, this object deserves continued observational and theoretical attention as the youngest known eclipsing binary system. 
  
\end{abstract}
 
 
\section{Introduction}

$\theta^{1}$ Orionis B, also known by its variable star designation of BM Ori, is the faintest of the {\it Trapezium} stars at the heart of the Orion Nebula. High resolution imagery resolves it into four distinct components, designated B$_1$ - B$_4$, all within about 1\arcsec\ of each other and likely interacting gravitationally \citep{m94,s99,c03,c12}. The brightest of these, $\theta^{1}$ Orionis B$_1$, is an eclipsing binary with a period close to 6.5 days \citep{h22,s44}. The five members of this small cluster or sub-group, totaling $\sim$15 M$_\odot$ within a volume of radius $\sim$500 AU, have contraction ages of less than 10$^5$ yr \citep{p01} and a dynamical time scale of only a few thousand years. Exposure ages of the proplyds in Orion and other aspects of the nebula also suggest a time scale of $\sim$10$^4$ - $10^5$ yr for the turn-on of high mass stellar activity in this region \citep{o09}. Except for the eclipsing binary, this system has an unstable or marginal configuration that is not likely to last long \citep{c12}. $\theta^{1}$ Orionis B$_1$ is the youngest eclipsing system known, and the secondary is probably among the least evolved stars known with a mass near 2 M$_\odot$ \citep{p76,vetal96} and an age of 10$^4$ - 10$^5$ yr.

The eclipsing binary system, which we will now simply refer to as BM Ori, exhibits unique and enigmatic properties, heightening our interest in it. The primary is a B3 star and appears to be a rather normal ZAMS object, with a mass of $\sim$6 M$_\odot$ and a radius of $\sim$3 R$_\odot$ \citep{p76}. Recognized as an eclipsing binary for nearly a century \citep{h22}, the first definitive light curves for the object were obtained in the 1960's and 1970's and showed an eclipse duration of about 16 hours with a flat-bottomed shape, indicative of a total eclipse, that lasted 8 - 9 hours \citep{h69,a76}. This was inconsistent, however, with spectral results which showed only a very weak A - F type spectrum even during the deepest minimum, or what came to be called ``pseudototality'' \citep{d70,p76}. The D-lines of NaI were strong enough to reveal the orbital motion of the secondary, and its mass is about 1/3 that of the primary. To account for the pseudototality of the eclipse, the secondary must have a radius comparable or larger than the primary, and so by standard stellar evolutionary models its light should be easily detectable, if not brighter than the B3 star. 

\citet{h71} suggested a way out of this conundrum, proposing that the secondary was a highly flattened proto-star. \citet{h75} later carried out a more detailed model of the eclipse by an occulting, reflecting, and self-luminous ``thick disk" (a rectangle in projection). He also noted the resemblance of BM Ori's light curve to that of the well-known (but physically much different) system $\epsilon$ Aur. The disk orbiting $\epsilon$ Aur has now been imaged, lending credence to the thick disk interpretation \citep{k10}. While this model is non-physical and oversimplified, it provides a good fit to the observed light curve and a first order interpretation of the photometry. 

Despite BM Ori's potential importance to star formation studies, many features of its light curve remain uncertain or controversial. The observed data show that the system reddens during eclipse such that the eclipse depth is greater at U than at V \citep{h69,a76}. \citet{a76} found some evidence for secular variation in the depth and duration of totality. The slope of the ``flat" bottom seemed to change direction over the years, and there is a persistent appearance of irregular variations at the level of up to 10\%. ``Shoulders" in the light curve before and after eclipse seem to come and go \citep{w94, b95}. Some authors \citep{b97} even question the reality of a small secondary eclipse apparently visible in the light curve of \citet{h69}. These properties suggested that this binary system in formation is a potential site for some interesting physical processes and motivated additional observations \citep{s77, a89, v96, v00}.

It actually seems surprising, at first, that such a potentially important eclipsing binary situated in such a well-known star forming region has not attracted more observational attention over the years, but there are real challenges. While relatively bright, the system is situated in one of the most crowded and naturally light polluted regions of the sky. Consequently, single channel photometry done before the era of CCDs was difficult, requiring care in centering the star, correcting for background and avoiding scattered light from other stars. Of course, it was unknown before the early 1990's that the observed system was a composite of objects so all photometric and spectroscopic measurements made then (and since) refer to the combined light of the five components. In addition, the period and duration of the eclipse cause difficulties for ground-based observers confined to a single longitude. Eclipses occur during night hours only $\sim$1/13 of the time and, with a duration of 16 hours, it is never possible to observe a full eclipse during a single night. It generally takes at least two years or more to fill out most phases of the light curve.

In this paper, we report results from a continuous monitoring campaign spanning 27 days obtained with the MOST satellite \citep{w03}. This was part of a larger project to monitor variable stars in the Orion Nebula and details of the observing program and other results are presented by \citet{t12}. We also obtained useful optical data at Van Vleck Observatory (VVO) on the campus of Wesleyan University during the 2007 - 2008 observing season. In addition, we re-examined the 10 days of data obtained on the Orion cluster at X-ray wavelengths during the {\it Chandra} Orion Ultradeep Project (COUP) study \citep{g05} and demonstrate that one of the BM Ori primary eclipses is detected in those data. We apply the model of \citet{h75} to analyze the optical light curve and discuss the evolution that has occurred in the secondary. An extensive set of high resolution spectra of BM Ori has also been obtained, which will be reported upon and interpreted elsewhere \citep{m12}.  
  
\section{Observations}

\subsection{The MOST Satellite Data}

The Microvariability and Oscillations of Stars (MOST) satellite is a project of the Canadian Space Agency designed to detect and characterize acoustic oscillations in solar-like stars and search for exoplanets by the transit method. It is capable of high precision photometry on very bright (uncrowded) stars. The telescope mirror, only 15 cm in diameter, feeds a $1024 \times 1024$ CCD through a single broadband filter that has a relatively flat response from 350 - 700 nm, with a central wavelength of about 525 nm, closest to the V band of the Johnson system. The microsatellite was launched in 2003 and is described in detail by \citet{w03}.

Our observing run stretched from JD 24555544.26 (2010 December 13) to JD 24555571.42 (2011 January 9) spanning about 4 cycles of BM Ori's orbit. Coverage was more or less continuous during this time except for gaps of 25 minutes during each 91 minute orbit of the spacecraft. A full description of the data reduction process, which involved ``pre-whitening" to take out signals induced by various spacecraft motions is given by \citet{t12}. The final data product presented and used here is the pre-whitened data binned in segments of about fifty minutes to reduce noise.

As noted previously, the MOST light curve includes contributions from all of the components, B$_1$ through B$_5$, although certainly the light of B$_1$ dominates. In addition, it is important to keep in mind that the field in which the eclipsing binary is situated is very crowded; scattered light from the other {\it Trapezium} stars, particularly $\theta^{1}$ Orionis C, and from the nebula is a problem. The $\sim$0.88\arcsec\ spatial resolution (relatively poor for space observations) and $\sim$1\arcsec\ pointing variations of the telescope conspire to make the data no more precise than could be obtained from the ground. The averaged data points have a typical precision based on their repeatability of about 0.005 mag. However, the wealth of data and the continuity of the phase coverage over four cycles make up for this. Again, this is comparable to what can be done from the ground but one would require a suite of telescopes observing in coordinated fashion around the world to get the phase coverage of the space-based data. Accordingly, the light curve for BM Ori presented here is much better than has ever been constructed for this star previously. 

\subsection{Photometric Data from Van Vleck Observatory}

Photometric observations were carried out with a CCD camera attached to the 0.6 m Perkin Telescope at the Van Vleck Observatory on the campus of Wesleyan University in Middletown, CT during the 2007 - 2008 observing season. This was the first season of operation for a new camera, which consists of a thermoelectrically cooled Apogee Alta U42 back-illuminated CCD with a standard UBVRI filter set from Santa Barbara Instrument Group (SBIG). Dates of primary and secondary minima were especially targeted. Ten nights of usable data were obtained, including three primary eclipse nights and two secondary eclipse nights. This resulted in approximately 1300 data points and provided a robust data set.  

Very short exposure times, ranging from about 0.2 sec in I to about 8 sec in U were required to avoid saturating the very bright stars in the {\it Trapezium}. The standard ``petal-type" camera shutter imposes a structure on the flat field for exposures of about 4 sec or less. When flat fields could not be taken with short enough exposure times, a shutter correction was included during the flat-fielding process. The data were reduced using standard IRAF tasks including the aperture photometry routine {\it apphot}. A typical aperture diameter of about 5\arcsec\ was used for the photometry to maximize signal-to-noise ratio while minimizing contamination from companion stars. The local sky was estimated as the median value in an annulus of inner radius 7\arcsec\ and width 3.5\arcsec\ after elimination of bright values due to neighboring stars or cosmic rays. All magnitudes were obtained differentially with respect to the primary comparison star $\theta^{1}$ Orionis D. The constancy of the comparison star was checked by differential photometry with respect to $\theta^{2}$ Orionis B. Based on this, we estimate the accuracy of each data point to be about 0.02 mag. The data are given in Table 1. 

\subsection{The X-ray Light Curve from COUP}

In January of 2003, the {\it Chandra} Orion Ultradeep Project (COUP) obtained a series of observations totaling nearly ten days of exposure time targeting the stars of the Orion Nebula Cluster \citep{g05}. As part of this large study, \citet{COUP} presented an X-ray light curve for $\theta^{1}$ Orionis B$_1$ (Fig. 5 of their paper) and identified it as a variable object. Their designation is COUP 778 and they also refer to $\theta^{1}$ Orionis B$_1$ as $\theta^{1}$ Orionis BE to distinguish it from a second (and stronger) source COUP 776 that is positionally coincident with the extremely close pair $\theta^1$ Orionis B$_2$ and B$_3$. It is noteworthy that {\it Chandra} was able to resolve $\theta^1$ Orionis B at least into its eastern and western components. \citet{COUP} refer to the unresolved pair B$_2$ and B$_3$ (COUP 776) as BW. The BW source is about three times more luminous than the BE source (BM Ori), and the separation between the eastern and western components is only about 1\arcsec.

The COUP 778 (BM Ori) light curve is relatively flat, with the exceptions of an apparent flare just after $4 \times 10^{5}$ sec and two prominent troughs at $7 \times 10^{5}$ and $8 \times 10^{5}$ sec. One trough appears to be caused by the edges of a gap in the COUP data. The other, however, is significant at the 95\% level according to the Bayesian block analysis of that paper, and its timing and duration allow us to associate it with the primary eclipse in BM Ori, as discussed in the following paragraphs. The apparent flare is coincident with a flare in COUP 776, and the authors believe the apparent flare in BM Ori is simply contamination from that event. They cannot distinguish whether the flare occurred in component B$_2$ or B$_3$ but it was evidently not in BM Ori. 

We re-analyzed the COUP data to assess the reality of the eclipsing binary's variability under the presence of the nearby source COUP 776. We were unable to fully isolate the contribution from either star due to the extreme overlap of their PSFs. Therefore, we defined an aperture size corresponding to a spatial region for BM Ori that excluded as much of the PSF of COUP 776 as possible while still retaining enough counts for a statistically significant analysis. We defined a comparable spatial region for COUP 776, which is considerably brighter than BM Ori at X-ray wavelengths. 

We followed the procedure described in \citet{w05} to search for variability using maximum-likelihood blocks (MLBs), but increased the significance for change points from 95\% to 99\%. In agreement with \citet{COUP}, we also found a significant reduction in the X-ray flux coincident with the optical eclipse. The only other variable feature in our light curve is the flare attributed to COUP 776 by \citet{COUP}. To check the source of this flare, we performed the same MLB analysis on COUP 776 and found that the flare is its dominant variability component. Furthermore, there is no evidence for an X-ray flux reduction in COUP 776 at the time of the optical eclipse in BM Ori. As a final check, we analyzed the light curves of three randomly selected COUP sources with our methods and found no indication of a flux reduction at the time of the optical eclipse of BM Ori in any of them. Thus the flare seen in BM Ori is undoubtedly due to contamination from COUP 776, as \citet{COUP} proposed, whereas the flux reduction is isolated to BM Ori.

We also utilized the Bayesian Estimate of Hardness Ratios (BEHR) code described in Park et al. (2006) to determine the hardness ratios for each time block as determined by the MLB method. This Bayesian estimation is superior to simple algebraic computations of hardness ratio when in the low photon count (Poisson-limited) regime. While the X-ray spectrum does apparently harden when the flux decreases, as may be seen in Fig. 5 of \citet{COUP}, the significance of the variation was only slightly higher than 1$\sigma$ in our analysis. 

\section{Results}

\subsection{The MOST Light Curve}

\begin{figure}
\epsscale{1.0}
\plotone{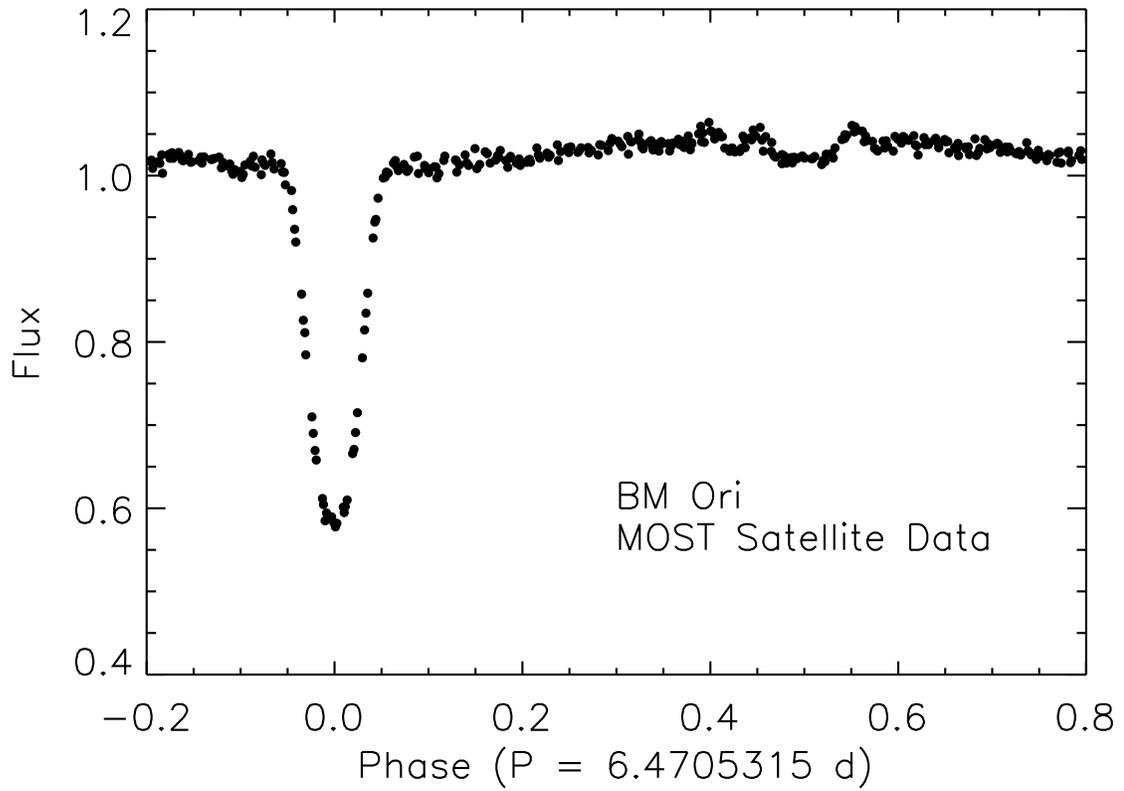}
\caption{Light curve of BM Ori obtained with the MOST satellite over four cycles phased with the orbital period. Features include the prominent primary eclipse at phase zero, an obvious secondary eclipse at phase 0.5, a clear reflection effect and some weak features shown better in Fig. 2.}
\label{MOST_light_curve}
\end{figure}

In Fig. \ref{MOST_light_curve} we show the MOST satellite flux measurements of BM Ori phased with its orbital period. The unfolded light curve, covering nearly four full cycles, showed no significant differences from cycle to cycle. The zero point for phase was determined from these data and a revised ephemeris is given in the next section. A broad ($\sim$0.1 in phase, or 16 hours) primary minimum is readily apparent, as is the equally broad secondary minimum centered at phase 0.5. In addition, the light curve exhibits a gradual rise in out-of-eclipse brightness from phase 0.1 to 0.4 and corresponding decline from phase 0.6 to 0.9 -- the so-called ``reflection effect" that is commonly observed in close binary systems. There is no evidence for any ``ellipsoidal variation," i.e., relative peaks near phase 0.25 and 0.75 caused by tidal distortion of the stars, as is seen in some close binary systems. In these respects the MOST light curve agrees with the best previous ground-based studies, done predominantly in the V magnitude system \citep{h69,a76}. 

A new feature of the light curve, indiscernible from the ground due to the larger uncertainties in the data, is the existence of some smaller brightness fluctuations that remained stable in phase over the four cycles monitored by MOST. They may be seen on Fig.~\ref{MOST_light_curve}, but are more obvious on Fig.~\ref{primary_secondary_comparison}. Here we have expanded the phase axis and overplotted the two halves of the light curve, one centered on primary eclipse and the other centered on secondary eclipse but shifted in phase by exactly one-half cycle. This plot demonstrates that the secondary eclipse is flat-bottomed and centered precisely on phase 0.5. There are also clear ``shoulders" associated with secondary eclipse, which occur exactly one-half cycle from when primary eclipse begins and ends. Furthermore, Fig.~\ref{MOST_light_curve} exhibits a small decrease in brightness at phase = -0.1 that appears to be paired with a brightness increase at phase = 0.4. A similar, but shallower and marginally significant ``matched pair" of deviations occurs close to phase = (0.1, 0.6). A simple explanation would be that the same material producing absorption when on the near side (phases = -0.1 and +0.1) appears in reflection on the far side (phases = 0.4 and 0.6), but this is probably overly simple as discussed later. Here we merely note the existence of variations in brightness in both directions which persist over at least four cycles and may occur in pairs separated by exactly one-half cycle.   

The breadth of primary eclipse raises the issue of whether all the transiting matter lies within the Roche lobe of the secondary. The duration (d) in phase of the transit of material confined within the Roche lobe of the secondary in a circular orbit may be written as
$$ d = {1 \over \pi} \sin^{-1} {(R_l + R_p) \over a} \ , $$
where R$_l$ is the radius of the Roche lobe of the secondary, approximated as a sphere, R$_p$ is the radius of the primary star, and $a$ is the separation of the stars. We estimate R$_l$ according to the formulation from \citet{e83}:
$$ R_l = \left({0.49 q^{2 \over 3} \over {{0.6 q^{2 \over 3} + \ln {(1 + q^{1 \over 3})}}}} \right) a $$ 
where $q$ is the ratio of secondary to primary mass. Using the radial velocity semi-amplitudes found by \citet{p76}, which have been recently confirmed by \citet{m12}, of K$_p$ = 52.8 km s$^{-1}$ and K$_s$ = 171 km s$^{-1}$, we determine q =  0.3088 and R$_l$/a $= 0.283 \pm 0.0145 $. Adopting R$_p$/a $= 0.085 \pm 0.002 $ (see Section 4), we find d$ = 0.120 \pm 0.005$.

As Fig.~\ref{primary_secondary_comparison} shows, we find that the bulk of the transiting material does lie within the Roche lobe of the secondary. Only the small, possibly paired features at phases (-0.1, 0.4) and (+0.1, 0.6) lie outside that limit. This development poses the interesting question of how the paired features could have survived in a phase plot if they are not within the Roche limit of the secondary and also not located at the L4 or L5 Lagrangian points (phases $\pm 0.17$). Non-periodic or transient features would not be so readily apparent in phase space folded to the binary period. We return to this intriguing result in Section 5.

\subsection{A Revised Ephemeris}

The time of minimum light for the epoch of the MOST observations can be estimated to within a minute or so using the symmetry of the light curve. It may be combined with times of minima reported over the decades, conveniently summarized by \citet{h71}. There are now ten epochs of minima spanning more than 5000 cycles. While the early estimates are more uncertain than recent ones, all of the data can be understood in terms of a simple ephemeris with no evidence for a change in period. We derive the following result, which is updated but entirely consistent with what was found by \citet{h71}:
$$ \rm{JD (min)} =  2455548.738 + 6.4705315\rm{E},$$
where epoch E is the number of periods. The dates are heliocentric and are used in all of the phase plots presented here. The error on the epoch of minimum is estimated to be $\pm 0.001$ d and the error on the period is estimated to be $\pm 1.5 \times 10^{-6}$ d.

\subsection{Comparison of the MOST Light Curve with VVO Data in V}

The partial light curves obtained at VVO allowed us to derive a depth of primary minimum of at least $0.57 \pm 0.01$ mag in V, which is consistent with previous studies. Color indices were also extracted for U-B, B-V, V-R, and R-I by combining differential magnitudes measured with respect to the primary comparison star. Note that we have not applied any transformation coefficients or second order extinction coefficients to these data so our quoted values for the colors may be slightly off, but this uncertainty is on the order of 0.01 mag and should not affect the slopes of the correlations with magnitude. The data indicate that the B-V color reddens by 0.06 mag during primary eclipse; this reddening is observed in the other bands as well. Table~\ref{table2} gives our measured values of the eclipse depths in the various filters, and they are quite consistent with what has been obtained by other authors \citep{h69,b95}. The depth of secondary eclipse is insufficient to determine what the color behavior of the system may be at this phase. 

The band pass of the MOST satellite CCD is broad, stretching from 350 to 700 nm, but its central wavelength of 525 nm is close to the central wavelength of the Johnson V band (550 nm). Since the 2010 - 2011 MOST results and 2007 - 2008 ground-based V data span similar epochs and center on similar wavelengths, it is reasonable to compare the two sets of light curves. We do this in Fig.~\ref{Full_Combined_light_curve}, focusing attention on primary eclipse. The green squares \citep{h69} and blue triangles \citep{a76} represent data from several decades ago, while the red x's (VVO) and black filled circles (MOST) represent the recent data presented in this paper. 

Clearly the shape of the primary eclipse has evolved dramatically over this period of time. The earliest light curve of sufficient precision available \citep{h69} shows the broadest minimum and the steepest slopes during ingress and egress. Data obtained a few years later \citep{a76} show excellent agreement with this shape but already an indication that the slopes may be decreasing and the minimum shortening in duration, as noted by the authors themselves. The more recent data, obtained at VVO in 2007 - 2008 clearly demonstrate that the duration of primary minimum, measured as the time between second and third contact, has shortened considerably, while its depth has not changed much, if at all. The MOST data, while obtained in a broader band pass than the others, match the recent ground-based V data quite well, and confirm that the duration of primary minimum has indeed shrunk, so much so that the minimum cannot now be accurately described as ``flat-bottomed" and it is not possible to identify points of second and third contact. These changes are indeed consistent with steady, secular evolution of the system, although we acknowledge that there is a substantial gap in observations between the epoch of \citet{a76} and modern times.  

Moreover, we lack reliable data at earlier epochs to further constrain what appears to be a significant secular evolution in the light curve. We do note, however, that \citet{s44} report that a 1940 edition of the {\it Katalog und Ephemeriden Ver\"anderlicher Sterne} \citep{s41} states that the duration of the eclipse is 23.5 hours, corresponding to 0.15 in phase, significantly longer than the 0.1 in phase characteristic of the modern epochs. This presumably refers to the duration from first to fourth contact, so it is technically not an extension of the trend we detail here but is a general indication that eighty or ninety years ago there was a larger degree of attenuation of the light of the primary during each orbital cycle than we observe today. The source also notes that the depth of the primary eclipse at those times was 0.61 mag, not significantly different from what is observed today in V (see Table~\ref{table2}). We return to the evolution of the light curve and what it implies for the physical nature of the system in the modeling section of this paper. 

\subsection{The X-ray Eclipse}

In Fig.~\ref{fig4} we show the COUP light curve of BM Ori phased with the optical ephemeris and overplotted with the MOST light curve. The X-ray eclipse is characterized by a flux reduction of about a factor of 3, which corresponds to 1.2 mag; this is deeper compared to the optical eclipse depths of $\lesssim$ 0.8 mag (see Table~\ref{table2}). The large uncertainties in the COUP data, however, make errors on the eclipse depth measurement $\sim$ 0.1 mag. There is a weak indication that the X-ray spectrum hardens during the eclipse (see Fig. 5 of \citet{COUP}), but this is only a 1$\sigma$ result in our analysis.

We removed 0.6 d of data centered on the flare in COUP 776 in forming this light curve. Otherwise, its contamination of the BM Ori light curve would have been evident in Fig. \ref{fig4}. Although the COUP study extended over 10 days and included two primary eclipses, only one contributes in the figure because the other happened to occur during a gap in the COUP record. Our analysis shows that there are no other significant features visible in the X-ray light curve other than primary eclipse. The absence of more subtle, i.e., non-primary eclipse, features may be physical in nature or merely an observational limit due to low signal-to-noise. We may conclude with some certainty, however, that the primary source of X-rays in the $\theta_1$ Orionis B system is the B3 star, not the secondary object of BM Ori, and that the X-ray emission arises close to the photosphere of the B star. 

It is interesting that the X-ray eclipse is, if anything, deeper than the optical eclipse. Two possible explanations are that: 1) the outer regions of the secondary are more opaque to X-rays than to optical light, making the projected area of the eclipsing object larger in the X-rays, or 2) there is a significantly larger contribution to the total light from either the secondary or from the other components of the system (B$_2$ and/or B$_3$) in the optical than at X-ray wavelengths. The latter possibility seems likely to be important since the depth of the eclipse increases to shorter wavelengths in the optical, where the secondary, B$_2$, and B$_3$ would contribute less. {\it Chandra} also resolved the sources and we were able to exclude much of the X-ray emission of COUP 776, while MOST (and all optical studies to date) did not resolve the system and the optical photometry, therefore, refers to the combined light of all components.  

\section{Modeling the Light Curve}
\setcounter{footnote}{2}
We employ a semi-empirical model developed by \citet{h75} to investigate the luminosity and dimensions of a binary system in which an obscured stellar component orbits the primary star. Huang treats the obscuring source as a geometrically thick disk around the secondary and considers two different scenarios regarding the contributions to the disk luminosity. In case A, the disk luminosity is due solely to scattered light from the obscured secondary; in case B, the disk luminosity stems from a combination of scattered light from the primary and that of the secondary. Given that the observed light curve exhibits clear reflection effects between phases, we adopt case B and compute the integrated light of the system accordingly using Simpson's rule. We refer the reader to the \citet{h75} paper for details of the thick disk model$^\thefootnote$. 

\footnotetext{Note that there is a typo in equation 7 of \citet{h75} concerning the value for $b$, which determines the varying surface brightness contribution of the disk edge. We derive analytically the numerator of $b$ to be $15(1+\beta')d$ instead of $5(1+\beta')d$; this correction was also confirmed numerically.}

We seek to understand changes in the light curve shape in the context of the thick disk model. For this reason, our main goal in modeling the modern observations is to find which parameters must be modified to accommodate for changes in the light curve behavior. As Figures~\ref{BMori_plot_full} and \ref{BMori_plot_zoom} clearly demonstrate, the published best-fit values from \citet{h75}, used to model the \citet{h69} light curve, no longer describe the system in the contemporary epoch. 

In particular, there are six features in the MOST light curve which differ from that of \citet{h69}: 
\vspace{-10pt}
\begin{enumerate} \itemsep1pt \parskip0pt 
\parsep0pt
	\item The slopes of ingress and egress are more shallow, 
	\item The duration of eclipse from second to third contact is shorter, 
	\item The reflection evident in phases between eclipses is less steep, 
	\item The bottom of the primary eclipse does not exhibit strong asymmetry,
	\item The secondary eclipse occurring at half-phase is better sampled, and 
	\item Tentative, shallow absorption and/or emission features are present at (-0.1, 0.4) and (0.1, 0.6). 
\end{enumerate}
\vspace{-10pt}
The model has 14 free parameters describing the orbital motion $\{r_1, r'_2, r''_2, \triangle, i, j, \Omega\}$ and the flux $\{l_1,\beta,\beta',\mu_1,\mu_2,c,d\}$ of a thick disk in circular motion around a stellar primary. Physically, the opacity may vary continuously with disk radius, but for simplicity we discretize the disk into two parts, an inner region ($0 \le r \le r'_2$) and an outer region ($r'_2 < r \le r''_2$). The free parameters are defined and tabulated in Table~\ref{tab1}. We also report our best-fit values, which match the MOST data quite well, in Table~\ref{tab1} (also see Figures~\ref{BMori_plot_full} and \ref{BMori_plot_zoom}). We used Huang's best-fit solutions as initial values for modeling the MOST light curve. We then varied each of the parameters individually to explore its effect(s) on the light curve shape with respect to the five features mentioned above. We describe our process below. 

Increasing the limb darkening values of the primary and the disk ($\beta$ and $\beta'$, respectively) produces only marginally deeper primary eclipses ($< 0.01$ in flux) and no perceivable changes in egress and ingress slopes. Furthermore, limb darkening effects are less severe in the blue, where the primary B3 star peaks, than red. For this reason we adopt Huang's values for $\beta$ and $\beta'$. $l_1$, the flux of the primary, is scaled such that the total flux of the system, i.e., primary star and disk, matches the observed maximum light level. Since the primary is a normal ZAMS star, its radius $r_1$ should not change on half-century time scales, and from the constancy of the total duration of primary eclipse, we hold $r_1$ fixed and put tight constraints on $r_2''$. As Huang himself points out, strong degeneracies exist between the physical thickness and inclination of the disk, since changing either would modify the projected disk height and thus the shape of the primary eclipse. For this reason, lower boundaries on $i$ and $j$ ($> 80^{\circ}$) are placed in order to determine an upper limit on the half-thickness of the obscuring disk $\triangle$. The best-fit $\triangle$ value derived in Table \ref{tab1} is then the disk half-height in the limit that projection effects are minimal. The third inclination, $\Omega$, describes the angle between the line of nodes of the orbit and that of the disk plane. We match feature 4 by changing $\Omega$ from $3^{\circ}$ to $ 0^{\circ}$; we note, however, that the scatter and shortened duration of the minimum do not allow us to well constrain the angle, and so this result is not significant. 

Since we do not consider changes in the sizes of the primary and secondary components for reasons enumerated above, features 1 and 2 are most readily achieved by decreasing $\mu_2$, which corresponds to the opacity of the outer region of the thick disk. Shallower slopes between primary and secondary eclipses indicate changes in $c$ and $d$, the constant and varying contributions of reflected light from the disk, respectively. We find that decreasing $c$ and $d$ values matches feature 4, as well as the depths of primary and secondary (feature 5) eclipses. Finally, we treat feature 6 in two manners. In case I, we treat the absorption features as simply scatter in the reflective portion of the light curve. In case II, we assume the absorption features are real and thus fit a steeper value to the reflection component. See Figures~\ref{BMori_plot_full} and \ref{BMori_plot_zoom} for a comparison between the data and different model light curves. 

We note that because the MOST data is broadband (3500 - 7000 \AA) whereas previous data in literature are in UBVRI filters \citep{h69, w69, a76, b95}, we cannot conclude from MOST data alone that evolution of the system is responsible for all of the light curve changes found. However, the recent UBVRI data obtained at VVO show that features 1, 2, and 4 are also present in V band and thus real. Furthermore, the difference in primary eclipse depth between MOST broadband data and previous V-band data is negligible, i.e., $<0.005$ in flux, whereas changes in $d$ values between case I (case II) and \citet{h75}'s produce reflection depth differences of $\sim0.025$ ($\sim0.01$) in flux. This indicates that the broadband vs. V band instrumental differences do not fully address the change in slope in the reflective portion of the light curve. The features 1, 2, and 3 are consistent with a decrease in opacity in the outer ring of the gaseous secondary disk, which we identify as the primary cause of the evolution of the light curve in the context of the thick disk model. As noted below, this is not a unique solution, but provides a possible framework to consider the changes in BM Ori's light curve. 

\section{Discussion}

BM Ori is among the youngest stellar objects known and the dramatic changes in its light curve reported here suggest that it has recently undergone a drop in the opacity of gas associated with its disk-shaped secondary. We turn to models of accreting binary systems for guidance in interpreting these observations, while cautioning that there is no direct evidence that BM Ori has a circumbinary (CB) disk or is still in its accretion phase. \citet{d11} have recently modeled accretion in close binary systems including those of unequal mass in circular orbits, similar to BM Ori. They show that quasi-stationary gas flows arise in the rotating frame of such systems with spiral paths connecting the secondary, and sometimes the primary, to the CB disk (see their Fig. 2). Circumstellar (CS) disks form around both stars, within their respective Roche lobes, and serve as a reservoir for stellar accretion.  

A schematic representation of how this might look in the BM Ori system is shown in Fig. \ref{model}. It is based on Fig. 10 of \citet{d11}, which is for a 5:1 mass ratio -- the closest to the BM Ori case shown in their paper. We have only shown the two densest gas streams connecting the stars to the CB disk. The one entering through the Lagrangian point L2 is densest and falls directly onto the secondary through the leading edge of its Roche lobe. The other, entering through L3 also falls on the secondary for this mass ratio and is of lower density. It feeds the circumstellar disk of the secondary through the trailing edge of its Roche lobe. In this model the highest density of gas, by far, is within the CS disk surrounding the secondary. The authors do not model its inevitable accretion onto the secondary.

It is beyond the scope of this paper to attempt a quantitative fit between BM Ori and a model such as the ones \citet{d11} calculate, but we do note some qualitative agreement. First, the accretion model provides a framework for understanding the basic structure of the system -- namely, a disk-like secondary orbiting a ZAMS primary. At this point in its evolution, most of the accretion is onto the secondary and the existence of a CS disk around the secondary is a natural product of that. Such a picture also accounts for the main discovery reported here -- that the primary eclipse has dramatically changed its shape in a way that can be modeled by an opacity change in the CS disk of the secondary. Random fluctuations or, perhaps, an overall decline in the accretion rate as the CB disk is depleted, could presumably account for the observed decline in opacity within the CS disk, which may at times drain faster than it is replenished.       

\begin{figure}
\epsscale{1.0}
\plotone{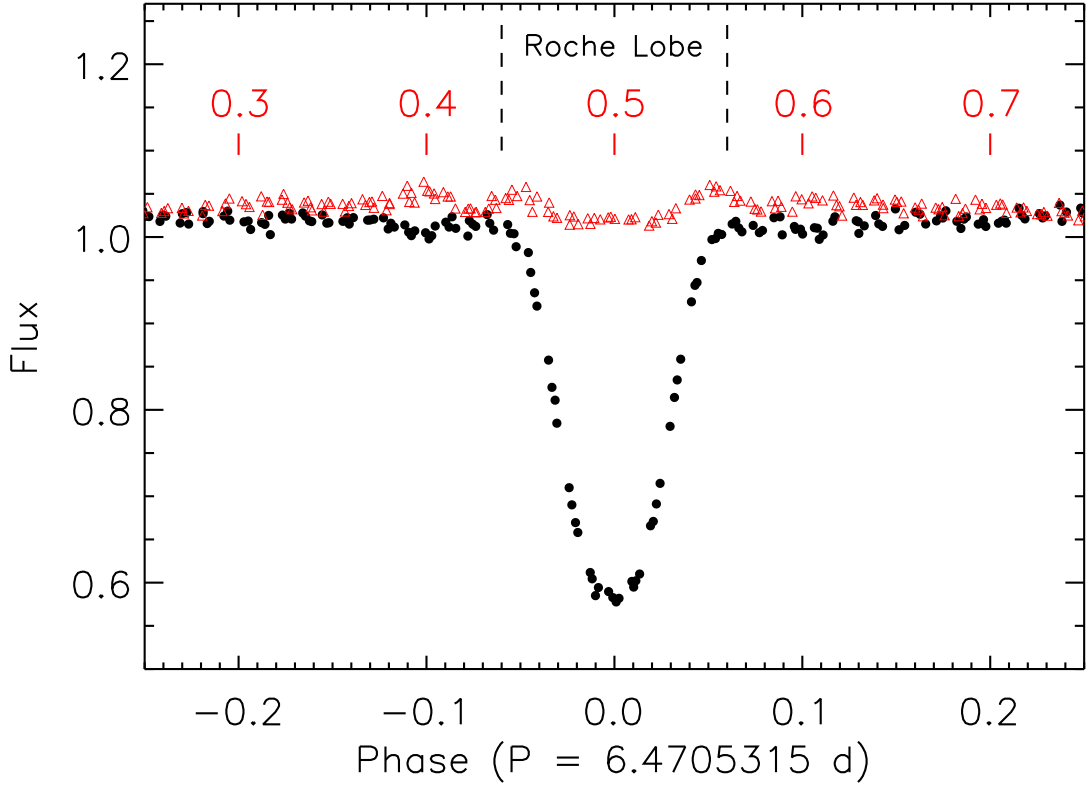}
\caption{Same as Fig. 1 except here we plot the half cycle centered on primary eclipse in black and the half cycle centered on secondary eclipse in red. The latter cycle was shifted by exactly 0.5 phase in order to plot them on the same figure and for easy comparison. The secondary eclipse occurs exactly at phase 0.5. The reflection effect causes the red points to lie above the black ones outside of eclipse. Error bars are roughly the size of the points ($\sim$0.005 in flux). Increases in brightness (shoulders) on secondary eclipse are clearly seen at phases, one-half cycle later, corresponding to the beginning of primary eclipse. A decline in brightness at phase=-0.1 appears to be accompanied by a small peak in brightness exactly one-half cycle later. A similar, but less obvious paired event may be seen at phases near 0.1 and 0.6. The phase extent of the Roche lobe of the secondary is indicated.}
\label{primary_secondary_comparison}
\end{figure}

\begin{figure}
\epsscale{1.0}
\plotone{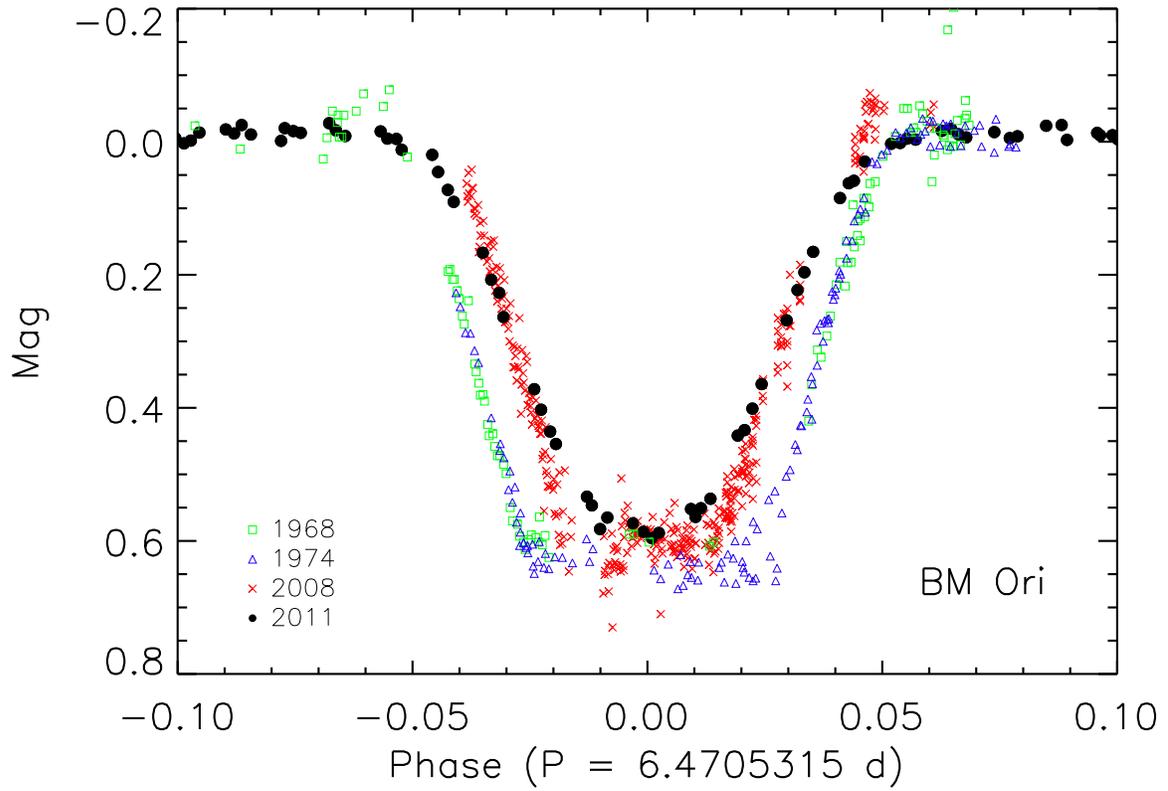}
\caption{The evolution of the shape of primary minimum over the years. Green squares are from \citet{h69}, blue triangles from \citet{a76}, red x's from VVO data obtained in 2007 - 2008 (see Table 1) and solid black circles are the 2010 - 2011 MOST data.}
\label{Full_Combined_light_curve}
\end{figure}

\begin{figure}
\epsscale{1.0}
\plotone{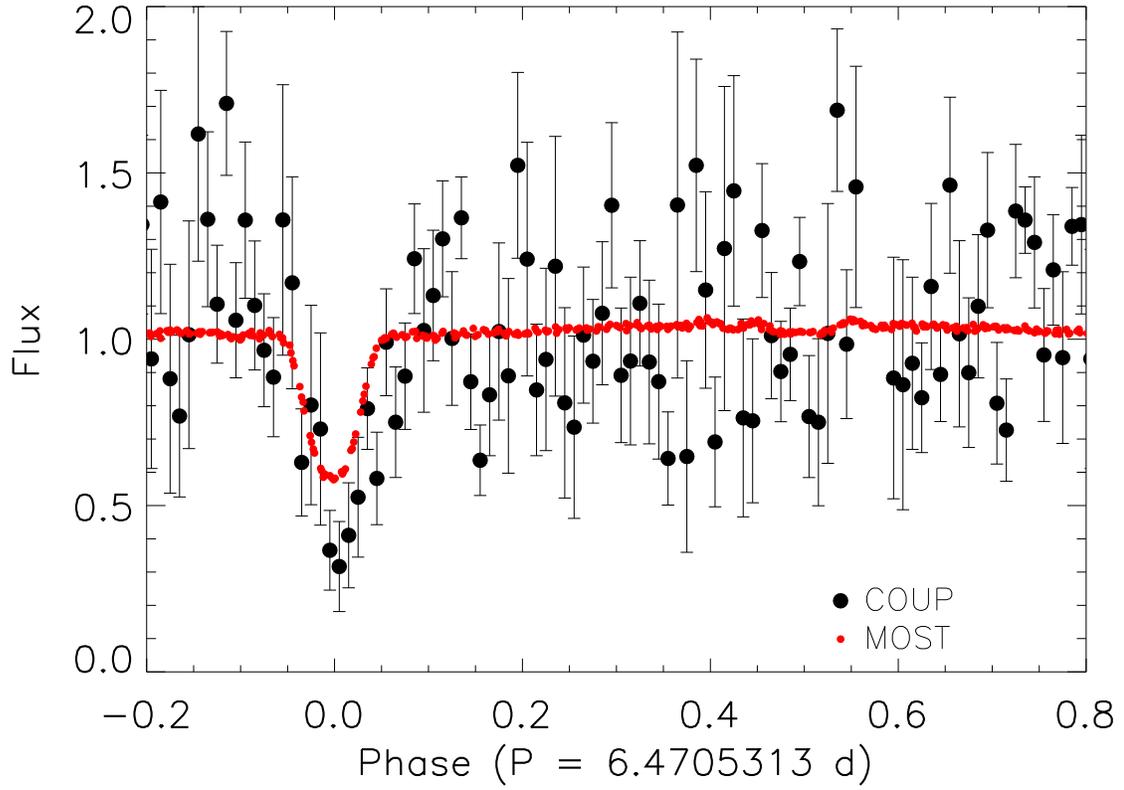}
\caption{X-ray data from the COUP survey \citep{g05} phased with the ephemeris of BM Ori. We clearly detect primary eclipse, coincident with the optical eclipse, in the COUP data. A flux of 1.0 corresponds to a COUP net count rate of $3.066 \times 10^{-3}$ cts s$^{-1}$. The COUP data were averaged into bins of 0.01 width in phase and the error bars are 1$\sigma$ errors of the mean.}
\label{fig4}
\end{figure}

\begin{figure}
\includegraphics[angle=90,scale=0.75]{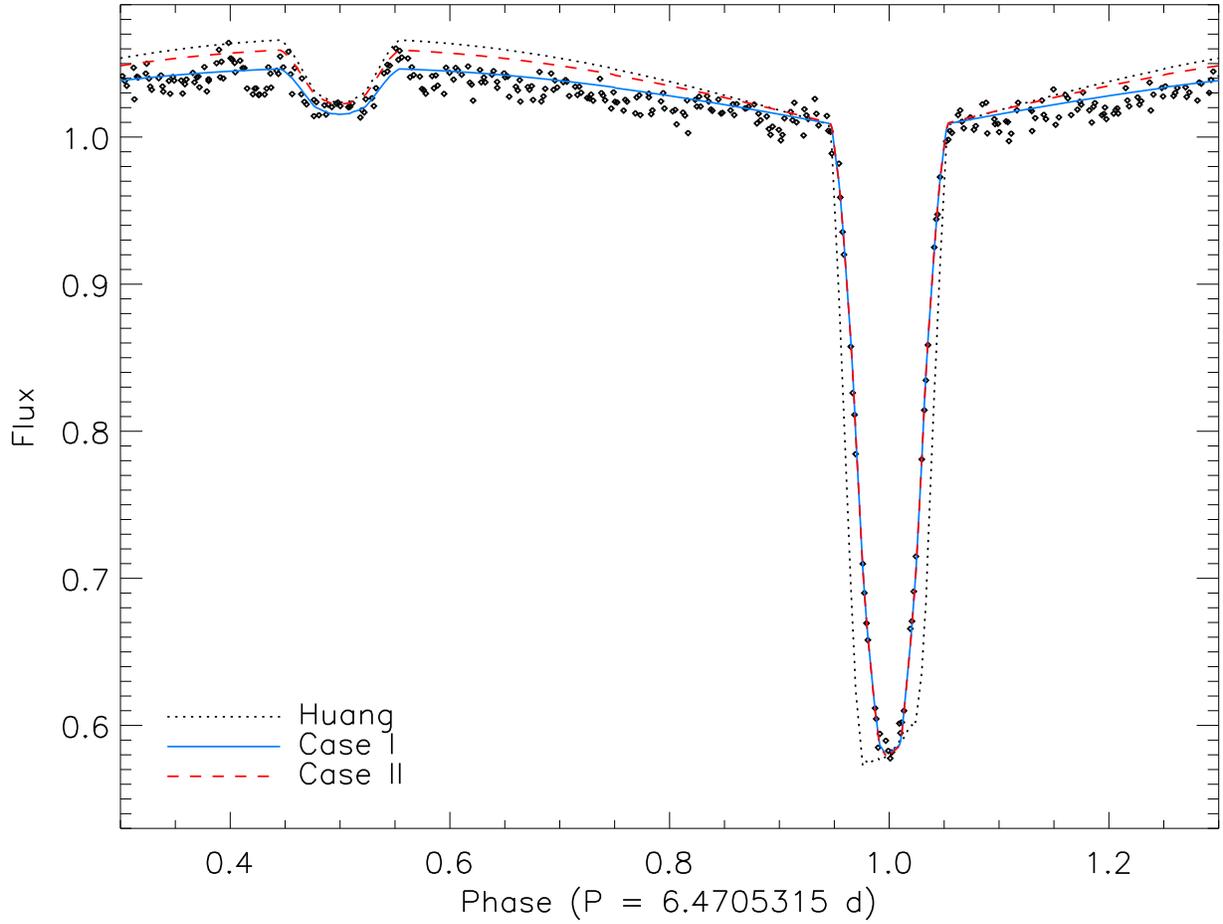}
\caption{Three different models overplotted against the MOST data where the dotted, solid, and dashed lines represent best-fits from \citet{h75}, case I, and case II, respectively. This figure shows observed data and the model fits in full. It clearly shows that Huang's fit no longer match the modern observations and demonstrates changes in the shape of the light curve, as discussed in the paper.  \label{BMori_plot_full}}
\end{figure}

\begin{figure}
\includegraphics[angle=90,scale=0.75]{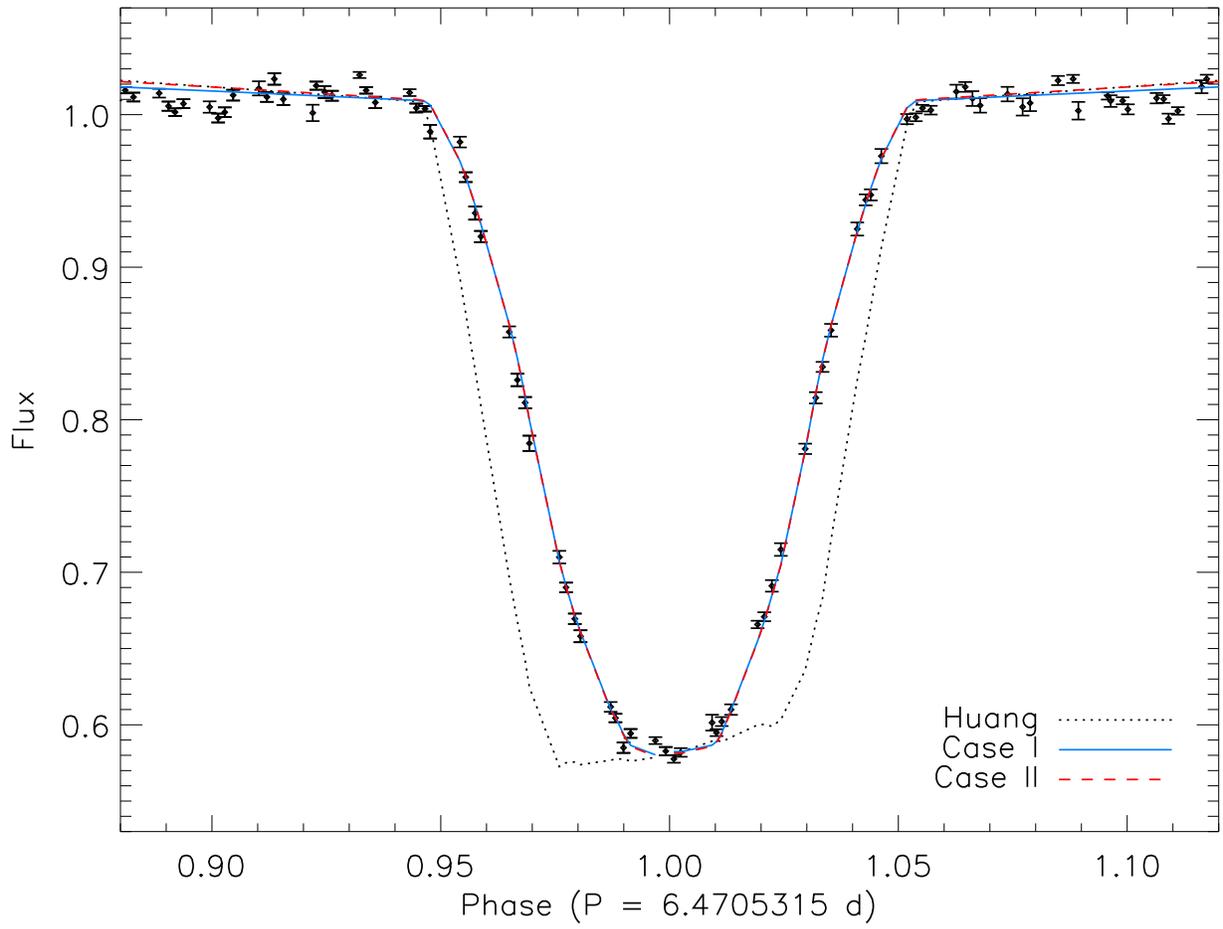}
\caption{A zoomed in version of Fig.~\ref{BMori_plot_full} showing the primary eclipse in detail. Error bars shown here are typical of all data points. \label{BMori_plot_zoom}}
\end{figure}

\begin{figure}
\epsscale{0.6}
\includegraphics[angle=90,scale=0.6]{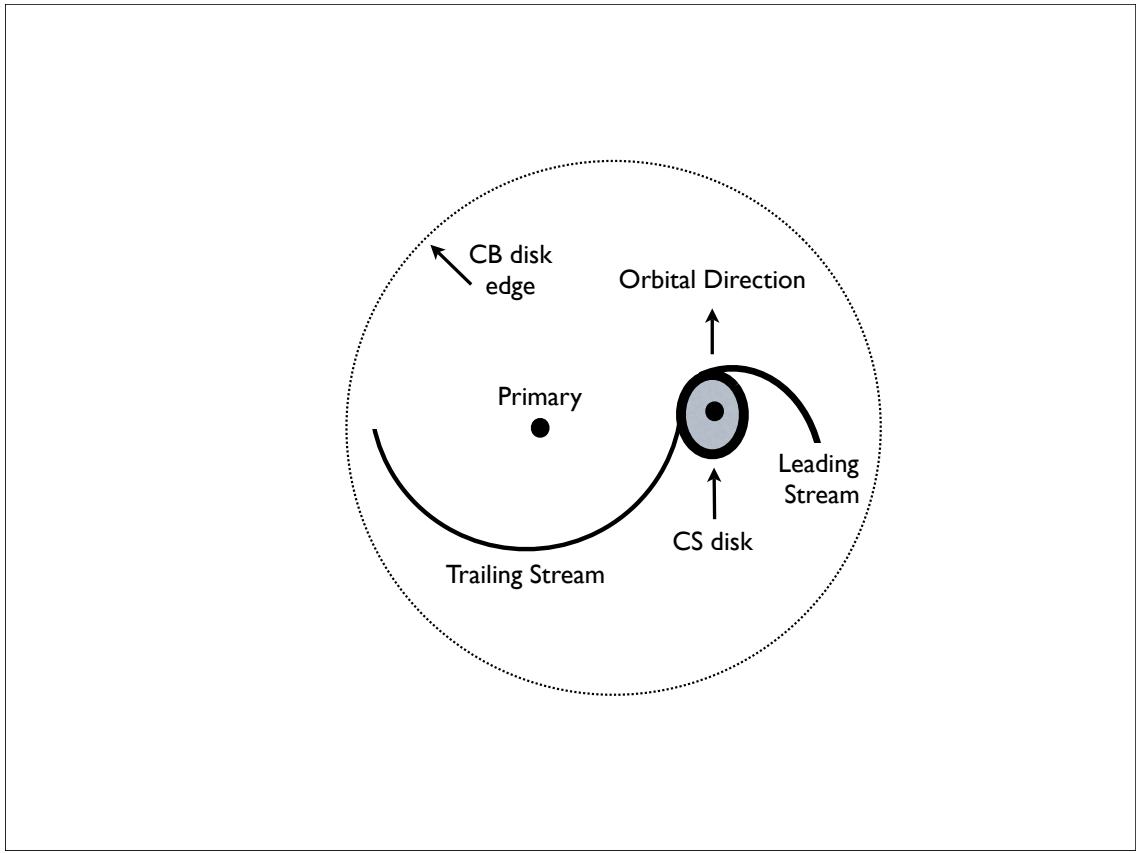}
\caption{A schematic diagram of what the BM Ori system may look like based on Fig. 10 of \citet{d11}. The location of the B3 primary star and CS disk surrounding the secondary are shown, as is the approximate inner edge of the CB disk. The approximate paths of the two major gas streams linking the CB disk to the CS disk are shown. The streams begin near the L2 and L3 Lagrangian points of the system. See text for discussion and caveats related to this model. \label{model}}
\end{figure}

The small, potentially paired features revealed in Fig. \ref{primary_secondary_comparison} at phases (-0.1,0.4) and (0.1,0.6) are intriguing in the context of this model. On the surface they seem to be concentrations of matter outside the Roche lobe of the secondary, somewhat beyond the current leading and trailing edges of the CS disk, that are seen in absorption when on the near side of the primary and emission on the far side. It is tempting to associate them with the quasi-stationary gas streams predicted in the \citet{d11} models. They even have the characteristic that the leading one is more prominent than the trailing one, as the model apparently predicts. It is difficult to understand how there could be any features at all outside the Roche lobe of the secondary (and not located at L4 or L5) orbiting with the binary period unless something like the quasi-stationary gas flows of the \citet{d11} model are invoked. That the model predicts roughly their locations and relative strengths is noteworthy.  

We caution, however, that while the pairing of these features on Fig. \ref{primary_secondary_comparison} is suggestive it may also be simply a coincidence. As noted in Section 4, the data can also be modeled with the assumption that the apparent emission features are actually just holes in the absorption. This avoids real difficulties in understanding how an object could appear as absorption on one side of the orbit and emission on the other in just the way required to match the observations. In particular, the material causing an eclipse on the near side would, when on the far side, be eclipsed itself; consequently it cannot be precisely the same matter. Perhaps it is an extension of that matter beyond the orbital plane of the secondary. Even then, however, the markedly peaked emission suggests a strongly asymmetric phase function for the reflection, one that might be associated with solids, but seems inappropriate to gaseous matter. Resolution of these issues will have to await more information about these features, including whether they are a permanent feature of the light curve. 

We have shown that the basic model of this system first proposed by \citet{h71} -- that the secondary is an elongated, perhaps disk-shaped structure -- continues to decently account for the main features of the light curve and its evolution. Since there has been dramatic secular evolution over the last few decades in the shape of the eclipse, it is also clear that there must be secular evolution in the size, shape, orientation and/or opacity of the occulting object in the past 40 years. We direct our analysis in the context of the thick disk model, which restricts the orientation and size of the system. We have shown that some relatively small changes to the model of \citet{h75} allow us to fit the modern light curve with the same basic model as he employed to fit the 1968 epoch light curve. The nature of these changes is to reduce the amount of obscuring and reflecting material composing the outer parts of the disk-shaped secondary. It is reasonable to suppose that this material is gradually accreting onto the secondary, which seems likely to be a star at a very early stage of its evolution.

Given the mass, size and likely temperature of the secondary, we suppose that the dominant opacity sources responsible for the eclipse in the optical would be Paaschen continuum absorption and, perhaps, electron scattering. The expected temperature at the inferred distance from the primary is already several thousand Kelvin. Moreover, illumination from the secondary component, which appears to have the temperature of an A or F star, contributes to heating of the disk material. Additional heating may come from the X-ray emission and the expected wind of the B star \citep{h75}. The environment appears much too hot and harsh for solid grains to survive in any form, so we presume the opacity is from the gas and, again, presume that H would dominate, as it does in stars of that temperature. 

If there is no substantial dust opacity associated with the secondary, then we need also to account for the clear reddening that occurs in the system light during primary eclipse. Again, it is beyond the scope of this paper to investigate that in detail, but we do mention that the expected contribution from components B$_2$ to B$_5$ needs to be taken into account. In this regard it is interesting that the X-ray eclipse is as deep, if not deeper, than the optical eclipse. The most likely explanation, in our view, is that visible light from the other components of $\theta^1$ Ori B is contributing substantially to the total system light and is more prominent than the X-rays during primary eclipse. 

Progress in understanding this system will require high resolution spectroscopy at all orbital phases and the creation of a physical model of the system. The spectroscopic study is well underway, with good phase coverage having been obtained using the HET spectrograph at a resolution of around 20,000. The spectra have been reduced and are currently being analyzed \citep{m12}. An improved physical picture of the system will undoubtedly come from that work and, hopefully, also by the efforts of groups devoted to understanding the formation of stars, particularly binary systems. As the youngest, least evolved example of such a system, displaying dramatic evolution on a time scale of years or decades, BM Ori undoutbedly has much yet to tell us about this process. Finally, we note that continued photometric monitoring from the ground, using a suite of telescopes at different longitudes, would be very worthwhile in assessing the permanence, or lack thereof, of the more subtle features in the light curve.

\acknowledgments

We thank Prof. Seth Redfield and Raquel Martinez of Wesleyan University for useful discussions on this object. MT would also like to thank all participants in the AAVSO's observing campaign on the {\it Trapezium}, whose work was very useful in the interpretation of the MOST photometry. This work was partially funded by the {\it NASA ROSES} program under grant NNX10AI83G and by NASA's Origins of Solar Systems program under a grant to WH. Student participation in this research was supported by an NSF/REU grant to Wesleyan University supporting the Keck Northeast Astronomy Consortium. We thank the referee for a careful reading and comprehensive report that improved our presentation of the material.

\clearpage

\begin{deluxetable}{ccccccc}
\tabletypesize{\scriptsize} \tablecaption{Photometry of BM Ori at Van Vleck Observatory \label{data}} \tablewidth{0pt} \tablehead { \colhead{Julian Date\tablenotemark{a}}   & \colhead{V} & \colhead{U-B} & \colhead{B-V} & \colhead{V-R} & \colhead{R-I} 
& \colhead{N\tablenotemark{b}}}

\startdata

2454403.8456  & 7.99 & -0.63 &  0.13 &  0.42 &  0.35 &  5 \\
2454403.8996  & 8.01 & -0.66 &  0.15 &  0.40 &  0.37 &  5 \\
2454484.5988  & 7.98 & -0.65 &  0.19 &  0.41 &  0.37 & 28 \\
2454487.5438  & 8.60 & -0.58 &  0.21 &  0.51 &  0.43 &  3 \\
2454487.5740  & 8.52 & -0.62 &  0.23 &  0.46 &  0.44 &  8 \\
2454487.5901  & 8.52 & -0.63 &  0.21 &  0.48 &  0.43 &  7 \\
2454487.6073  & 8.57 & -0.63 &  0.24 &  0.49 &  0.45 &  4 \\
2454487.6206  & 8.56 & -0.60 &  0.23 &  0.49 &  0.44 &  8 \\
2454487.6326  & 8.57 & -0.63 &  0.25 &  0.48 &  0.45 &  7 \\
2454489.6382  & 7.94 & -0.64 &  0.16 &  0.41 &  0.35 &  8 \\
2454496.5859  & 7.97 & -0.60 &  0.14 &  0.42 &  0.36 &  6 \\
2454497.5880  & 7.94 & -0.64 &  0.17 &  0.41 &  0.35 & 25 \\
2454500.5447  & 8.57 & -0.59 &  0.23 &  0.52 &  0.39 &  5 \\
2454500.5490  & 8.56 & -0.58 &  0.21 &  0.46 &  0.46 &  5 \\
2454500.5550  & 8.57 & -0.59 &  0.24 &  0.49 &  0.45 &  5 \\
2454500.5592  & 8.49 & -0.59 &  0.24 &  0.39 &  0.45 &  5 \\
2454500.5644  & 8.55 & -0.63 &  0.20 &  0.47 &  0.45 &  5 \\
2454513.5426  & 8.42 & -0.65 &  0.22 &  0.44 &  0.41 & 13 \\
2454513.5910  & 8.38 & -0.64 &  0.15 &  0.44 &  0.41 &  5 \\
2454513.6008  & 8.29 & -0.67 &  0.25 &  0.39 &  0.40 &  5 \\
2454516.5473  & 7.97 & -0.64 &  0.15 &  0.38 &  0.35 & 93 \\
\enddata

\tablenotetext{a}{Reported values are the mean Julian dates of N very short exposures; the precision is dominated by the readout time, $\sim$8 seconds.}
\tablenotetext{b}{N is the number of independent measurements averaged to form the quoted values.}

\end{deluxetable}

\clearpage

\begin{deluxetable}{cccc}
\tabletypesize{\scriptsize} 
\tablecaption{Eclipse Depths (mag) as a Function of Color}  

\tablewidth{0pt} 
\tablehead {\colhead{Filter}  & \colhead{HG\tablenotemark{a}}   & \colhead{BV\tablenotemark{b}} & \colhead{VVO\tablenotemark{c}}}

\startdata

U & 0.77 & 0.81 & 0.68 \\
B & 0.70 & 0.73 & 0.65 \\
V & 0.57 & 0.62 & 0.59 \\
R & - & 0.49 & 0.53 \\
I & - & 0.46 & 0.45 \\

\enddata

\tablenotetext{a}{\citet{h69}}
\tablenotetext{b}{\citet{b95}}
\tablenotetext{c}{This paper.}
\label{table2}
\end{deluxetable}

\clearpage

\begin{center}
\begin{deluxetable}{llllll}{}
\rotate
\tablewidth{533pt}
\tablecaption{Parameters for the Thick Disk Model}
\small
\tablehead{\multicolumn{1}{l}{} & \multicolumn{1}{l}{} & \multicolumn{1}{l}{} & \multicolumn{3}{c}{Current Paper}  \\
\multicolumn{1}{l}{Symbol} & \multicolumn{1}{l}{Significance} & \multicolumn{1}{l}{Huang (1975)} & \multicolumn{1}{l}{Case I} & \multicolumn{1}{l}{Case II} & \multicolumn{1}{l}{Uncertainty\tablenotemark{a}} }
\startdata
 $l_1$  & flux of primary star \tablenotemark{b} & $0.9$ & $0.967$ & $0.965$ & $0.002$ \\
 $r_1$  & radius of primary star & $0.085a$ & $0.085a$ & $0.085a$ & $0.001a$ \\
 $r_2'$  & inner radius of disk & \nodata & $0.13a$ & $0.13a$ & $0.01a$ \\
 $r_2''$  & outer radius of disk & $0.25a$ & $0.25a$ & $0.25a$ & $0.01a$ \\
 $\triangle$  & half-height of disk & $0.0282a$ & $0.0282a$ & $0.0282a$ & $0.002a$ \\
 $\beta$  & limb darkening constant for star & $2.0$ & $2.0$ & $2.0$ & $0.5$ \\
 $\beta'$  & limb darkening constant for disk & $2.0$ & $2.0$ & $2.0$ & $0.5$ \\
 $i$  & inclination of binary orbit w. r. t. observer & $89^{\circ}$ & $89^{\circ}$ & $89^{\circ}$ & $0.5^{\circ}$ \\
 $j$  & inclination of disk plane w. r. t. observer & $90^{\circ}$ & $90^{\circ}$ & $90^{\circ}$ & \nodata \\
 $\mu_1$  & absorption coefficient within $r_2'$ & $\infty$ & $\infty$ & $\infty$ & \nodata \\
 $\mu_2$  & absorption coefficient btw. $r_2'$ and $r_2''$ & $\infty$ & $2.4$ & $2.4$ & $0.1$ \\
 $c$  & constant disk light contribution & $0.04$ & $0.035$ & $0.035$ & $0.001$ \\
 $d$  & variable disk light contribution & $0.07$ & $0.045$ & $0.060$ & $0.001$ \\
 $\Omega$  & angle btw. line of nodes of orbit and of disk plane &$3^{\circ}$ & $0.0^{\circ}$ & $0.0^{\circ}$ & $3.0^{\circ}$ \\
\enddata

\tablenotetext{a}{Uncertainties are determined by varying each parameter from best-fit values while holding the rest fixed and visually inspecting when the fit no longer matched the observed data.}

\tablenotetext{b}{Fluxes are normalized such that the total flux of the model system ($l_1 + c + d$) corresponds to the maximum observed flux. }

\label{tab1}
\end{deluxetable}
\end{center}

\clearpage

\end{document}